%% file: main.tex
\documentclass[10pt,a4paper,twocolumn,superscriptaddress,aps,prd,longbibliography,nofootinbib]{revtex4-2}
\usepackage[modulo]{lineno}
\usepackage{amsmath}
\usepackage{mathtools}
\usepackage{amsfonts}
\usepackage{amssymb}
\usepackage{bbold}
\usepackage{graphicx}
\usepackage[usenames,dvipsnames]{color}
\usepackage{float}
\usepackage{multirow}
\usepackage{soul}
 
\usepackage{blindtext}

\begin{document}

\author{Jonas Erhardt}
\affiliation{Physikalisches Institut, Universit\"at W\"urzburg, D-97074 W\"urzburg, Germany}
\affiliation{W\"urzburg-Dresden Cluster of Excellence ct.qmat, Universit\"at W\"urzburg, D-97074 W\"urzburg, Germany}
 
\author{Mattia Iannetti}
\affiliation{Department of Physical and Chemical Sciences, University of L’Aquila, Via Vetoio, 67100, L’Aquila, Italy}

\author{F. Dominguez}
\affiliation{W\"urzburg-Dresden Cluster of Excellence ct.qmat, Universit\"at W\"urzburg, D-97074 W\"urzburg, Germany}
\affiliation{Institut f\"ur Theoretische Physik und Astrophysik, Universit\"at W\"urzburg, D-97074 W\"urzburg, Germany}

\author{Ewelina M. Hankiewicz}
\affiliation{W\"urzburg-Dresden Cluster of Excellence ct.qmat, Universit\"at W\"urzburg, D-97074 W\"urzburg, Germany}
\affiliation{Institut f\"ur Theoretische Physik und Astrophysik, Universit\"at W\"urzburg, D-97074 W\"urzburg, Germany}

\author{Bj\"orn Trauzettel}
\affiliation{W\"urzburg-Dresden Cluster of Excellence ct.qmat, Universit\"at W\"urzburg, D-97074 W\"urzburg, Germany}
\affiliation{Institut f\"ur Theoretische Physik und Astrophysik, Universit\"at W\"urzburg, D-97074 W\"urzburg, Germany}

\author{Gianni Profeta}
\affiliation{Department of Physical and Chemical Sciences, University of L’Aquila, Via Vetoio, 67100, L’Aquila, Italy}
\affiliation{SPIN-CNR, University of L’Aquila, Via Vetoio 10, 67100, L’Aquila, Italy}

\author{Domenico Di Sante}
\affiliation{Department of Physics and Astronomy, University of Bologna, Bologna, Italy}

\author{Giorgio Sangiovanni}
\affiliation{W\"urzburg-Dresden Cluster of Excellence ct.qmat, Universit\"at W\"urzburg, D-97074 W\"urzburg, Germany}
\affiliation{Institut f\"ur Theoretische Physik und Astrophysik, Universit\"at W\"urzburg, D-97074 W\"urzburg, Germany}

\author{Simon Moser}
\affiliation{Physikalisches Institut, Universit\"at W\"urzburg, D-97074 W\"urzburg, Germany}
\affiliation{W\"urzburg-Dresden Cluster of Excellence ct.qmat, Universit\"at W\"urzburg, D-97074 W\"urzburg, Germany}

\author{Ralph Claessen}
\email{e-mail: claessen@physik.uni-wuerzburg.de}
\affiliation{Physikalisches Institut, Universit\"at W\"urzburg, D-97074 W\"urzburg, Germany}
\affiliation{W\"urzburg-Dresden Cluster of Excellence ct.qmat, Universit\"at W\"urzburg, D-97074 W\"urzburg, Germany}

\date{\today}






\title{
Backscattering in Topological Edge States Despite Time-Reversal Symmetry
}
\maketitle


\bigskip
\noindent
{\bf

Spin-momentum-locked edge states of quantum spin Hall insulators (QSHIs) provide a compelling platform for spintronic applications, owing to their intrinsic protection against backscattering from non-magnetic disorder.
This protection emerges from time-reversal symmetry, which pairs Kramers partners of helical edge modes with opposite spin and momentum, thereby strictly forbidding elastic single-particle backscattering within the pair.
Yet, contrary to the idealized notion of linear edge bands, the non-monotonic dispersions of realistic materials can host multiple Kramers pairs, reintroducing backscattering channels between them without violating time-reversal symmetry.
Here, we investigate inter-Kramers pair backscattering in the non-linear edge bands of the QSHI indenene, highlighting a critical aspect of edge-state stability.
Using quasiparticle interference in scanning tunneling spectroscopy -- a direct probe of backscattering -- we observe pairwise coupling between energy-degenerate Kramers pairs, while energy regions with only a single Kramers pair remain strictly protected. 
Supported by theoretical analysis, our findings provide an unprecedented experimental demonstration of edge state backscattering fully consistent with their underlying topological protection. 
This insight has profound implications for numerous QSHI candidates, emphasizing that the mere presence of gap-traversing edge modes does not inherently guarantee their protection against backscattering.}

\input{TextBody}
\newpage
\input{Methods}


{
	We thank C. Li for helpful discussion. We are grateful for funding support from the Deutsche Forschungsgemeinschaft (DFG, German Research Foundation) under Germany’s Excellence Strategy through the Würzburg-Dresden Cluster of Excellence on Complexity and Topology in Quantum Matter ct.qmat (EXC 2147, Project ID 390858490) as well as through the Collaborative Research Center SFB 1170 ToCoTronics (Project ID 258499086).
    G.P. acknowledges financial support by the European Union – NextGenerationEU, Project code PE0000021 - CUP B53C22004060006 - “SUPERMOL”, “Network 4 Energy Sustainable Transition – NEST” and the European Union - NextGenerationEU under the Italian Ministry of University and Research (MUR) National Innovation Ecosystem grant ECS00000041 - VITALITY - CUP E13C22001060006.
}


\end{document}

%% file: TextBody.tex
\begin{figure}[b]
 \centering
 \includegraphics[width=1\columnwidth]{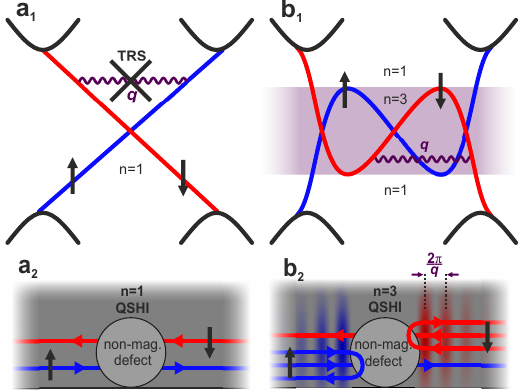}
\caption{\textbf{Backscattering between helical edge states.}
\textbf{a}$_1$ Schematic linear edge dispersion of a QSHI with one single Kramers pair within the topological bulk gap \cite{Kane2005a}. Blue and red colors mark channels of opposite spin polarizations. \textbf{a}$_2$ Corresponding real space cartoon of a right-moving spin up current (blue: positive group velocity $dE/dk$) and a left-moving ($dE/dk<0$) spin down current (red) allocated at a single edge of a QSHI. Scattering between these channels is prohibited by TRS.
\textbf{b}$_1$ Schematic non-linear but $\mathcal{S}$-shaped edge dispersion of a QSHI with energy regions of one and three Kramers pairs within the topological bulk gap, the latter indicated by purple shading. \textbf{b}$_2$ Real space cartoon of the three Kramers pairs scenario, with disparate numbers of left and right movers in the two spin channels. TRS-conserving backscattering between two Kramers pairs at non-magnetic defects produces standing waves with wavelength $2\pi/q$.
}
\label{Fig:1}
\end{figure}

Dissipationless ballistic transport in spin-polarized metallic edge modes of QSHIs forms the foundation of their technological potential \cite{Hasan2010,Spintronik1, Spintronik2, LowPower2, LowPower1, QuantenComp1} and sparked an intense search for new topological materials \cite{Bradlyn2017,Weber_2024}. 
In a QSHI, time-reversal symmetry (TRS) and the resulting spin-momentum locking protect Kramers partners -- time-reversal symmetric edge modes forming a Kramers pair -- from elastic single-particle backscattering.
This protection occurs because the momentum transfer $q$ required for backscattering would necessitate a TRS violating spin flip, which is therefore not allowed as indicated in Fig.~\ref{Fig:1}a$_1$ \cite{Xu2006}.
As a result, both modes are perfectly transmitted at non-magnetic defects and propagate freely along the edge when they are equally populated (Fig.~\ref{Fig:1}a$_2$).
In the non-equilibrium situation of electronic transport this would lead to a two terminal edge conductance of $2e^2/h$ \cite{Kane2005a}, where the factor of 2 accounts for the contributions from both edges.


%

Historically, this scenario was discussed mostly for linear dispersing edge bands of ideal honeycomb materials \cite{Kane2005a,Reis2017,Stuehler2022} and quantum wells \cite{Bernevig2006,Koenig2007,Knez2011} that produce only a \textit{single} Kramers pair (Fig.~\ref{Fig:1}a$_1$). However, this picture is often too simplistic given that in real QSHI materials a multi-orbital character \cite{Wang2014}, 
buckling \cite{Matthes2014}, 
or details in the edge potential \cite{Wang2017,Fu_2017} 
render the band dispersion of edge modes more complicated, typically highly non-linear \cite{Hasan2010,Qian2014,Zheng2016,Wu2018,Pulkin2020}.
Without repercussions on the bulk-boundary correspondence, such edge-specific modifications to the Hamiltonian can, in principle, produce an $\mathcal{S}$-shaped edge dispersion where an odd number $n=3$ of Kramers pairs within the bulk gap \textcolor{black}{are energetically degenerate (Fig.~\ref{Fig:1}b$_1$).}
In contrast to the pure linear band scenario of Fig.~\ref{Fig:1}a, the edge conductance in such systems should thus be either $2e^2/h$ or $6e^2/h$, depending on whether the precise position of the Fermi level activates $n=1$ or $n=3$ Kramers pairs (Fig.~\ref{Fig:1}b$_1$) \cite{Li2014,Bieniek2023}. 

However, while for the $n=1$ band section, TRS continues to strictly prohibit elastic scattering, 
the $n=3$ section contains right- and left moving modes of identical spin-character \textcolor{black}{that \textit{can} be connected by time reversal symmetric perturbations $V(q)$}, 
thus enabling single-particle backscattering \textit{between} different Kramers pairs (Fig.~\ref{Fig:1}b$_2$) \cite{Xu2006}. The interference between the incoming and the backscattered edge modes then leads to a pairwise elimination of Kramers pairs \cite{Xu2006}, and the two terminal edge conductance drops to $2e^2/h$ in the presence of disorder \cite{Hasan2010,Wada2011,Li2014,Bieniek2023,Martin1992,Fisher1981}.

Although many QSHI materials fall into the $n>1$ Kramers pair regime \cite{Wada2011,Zheng2016,Matthes2014,Fu_2017,Pulkin2020} and theoretical studies are extensive \cite{Xu2006,Wu2006,Kagalovsky2018,Li2014}, experimental reports of backscattering \textit{between} Kramers pairs remain rare \cite{Drozdov2014}, and in QSHI so far even missing. 
Here, we exploit the interference of incoming and backscattered edge-modes and their formation of standing waves that can be readily detected via quasi particle interference (QPI) in scanning tunneling spectroscopy (STS) (Fig.~\ref{Fig:1}b$_1$) \cite{Drozdov2014}. Investigating the energy dependent momentum transfer $q$ that characterizes the single-particle backscattering process, we experimentally confirm the pairwise elimination of Kramers pairs in the $\mathcal{S}$-shaped bandstructure of a QSHI with $n>1$, effectively restoring the $n=1$ regime with a single topologically protected Kramers pair \cite{Xu2006}.

\subsection{Bulk boundary correspondence in the QSHI indenene}

\begin{figure*}[t!]
 \centering
	\includegraphics[width=1\textwidth]{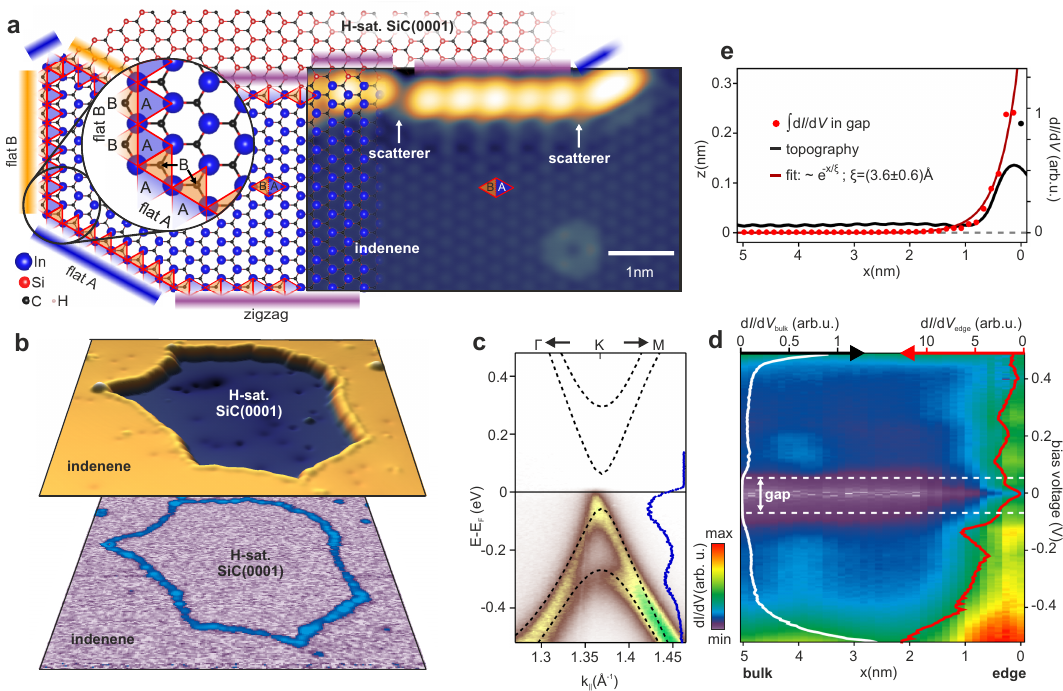}
\caption{\textbf{Bulk boundary correspondence in the QSHI indenene.}
\textbf{a} Ball and stick model (left) of indenene on SiC(0001) forming the three edge types (zigzag, flat A, flat B) at the boundary to H-saturated SiC. 
\textcolor{black}{The inset highlights the carbon position in the top SiC layer, distinguishing flat edge B from flat edge A.}
STM topography (right) of an indenene edge consisting of zigzag and flat edge A segments, interrupted by kinks that act as scattering centers (I$_{\text{T}}= 300 \,$pA, V$_{\text{bias}}= 0.95 \,$V). 
\textbf{b} Top: 3D visualisation of a (32$\,$nm$)^2$ STM topography scan (I$_{\text{T}}= 10 \,$pA, V$_{\text{bias}}= -3 \,$V) showing a hole with H-saturated SiC in an otherwise closed indenene film. Bottom: Corresponding $dI/dV$-map highlighting edge states and taken at constant height by integrating states within V$_{\text{bias}}= (0 \pm 10) \,$mV.
\textbf{c} ARPES measurement and overlaid \textcolor{black}{$G_0 W_0$ bandstructure (reprinted from \cite{Eck2022})} of indenene's Dirac bands, illustrating band position and charge neutral doping. The inset shows an energy distribution curve at the K-point of indenene.
\textbf{d} $dI/dV$ line-scan approaching an indenene zigzag edge, showing that metallic edge states (inset red curve) fill the 120$\,$mV bulk gap (inset white curve).
\textbf{e}
Edge topography (black) and exponential decay of metallic edge states with a decay constant of $\xi \, = (\,3.6 \pm 0.6$)\AA$ $ fitted to the gap-integrated $dI/dV$ signal (red dots).  
Tunneling parameters of \textbf{d,e} are I$_{\text{T}}= 50 \,$pA, V$_{\text{bias}}= 1 \,$V, $\delta$z$ = -0.2\,$nm (\textbf{e}).
}
\label{Fig:2}
\end{figure*}

The QSHI we chose for this study is indenene, a \textcolor{black}{triangular} $1\times 1$ indium monolayer grown epitaxially on SiC(0001), whose non-trivial topology has been recently demonstrated by independent bulk probes \cite{Bauernfeind2021, Erhardt2024}.
\textcolor{black}{By limiting the surface coverage in a controlled way to about 90$\%$ of a monolayer \cite{Erhardt2022}, we create holes in the film with clearly defined edge boundaries between indenene and the uncovered SiC surface. 
These boundaries exhibit three distinct indenene edge types, as illustrated on the left side of Fig.~\ref{Fig:2}a: flat A, flat B and zigzag.}

\textcolor{black}{While the zigzag edge bears a resemblance to edges of freestanding honeycomb lattices such as graphene, flat edges A and B terminate the indenene unit cell by either triangle A or B (see inset of Fig.~\ref{Fig:2}a ), which are inequivalent due to the carbon atom of the topmost SiC-layer positioned beneath triangle B, but not A \cite{Eck2022}.
}
In scanning tunneling microscopy (STM), they are differentiated by gauging the unit cell of indenene vs the position of characteristic subsurface defects, as described in more detail in Supplementary Discussion I. Typically, STM finds straight segments of alternating edge type, as exemplified in Fig.~\ref{Fig:2}a and at the top of Fig.~\ref{Fig:2}b. 
\textcolor{black}{They are separated by edge imperfections, such as kinks or defects (Fig.~\ref{Fig:2}a right), which serve as scattering centers yielding QPI patterns of edge modes.}

STS differential conductance (d$I$/d$V$) maps essentially reflect the local density of states (LDOS) and, when taken in a small energy interval around the Fermi level, reveal that the indenene edges are clearly metallic (Fig.~\ref{Fig:2}b, bottom). Monitoring the energy dependent LDOS across an indenene edge in Fig.~\ref{Fig:2}d, we find the bulk spectra to reproduce the topological energy gap of $E_{\text{gap}}\approx\,120\,$meV as previously found by ARPES \cite{Bauernfeind2021} and GW many-body pertubation theory \cite{Eck2022} in Fig.~\ref{Fig:2}c. Approaching the edge, this bulk gap is gradually filled with spectral weight, consistent with the bulk-boundary correspondence and the presence of topological edge modes (Fig.~\ref{Fig:2}b, bottom) \cite{Erhardt2024,Bauernfeind2021}.

Incidentally, we find the in-gap differential conductance to localize exponentially within a penetration depth $\xi \, = (\,3.6 \pm 0.6$)\AA$\, \,$ from the topographic edge, see Fig.~\ref{Fig:2}e and Supplementary Discussion II. \textcolor{black}{
Conversely, using the BHZ model's approximate relation yields $\xi\approx (dE/dk)/E_{\text{gap}}\approx 7$\;nm for edge bands hypothetically connecting valence and conduction bands linearly \cite{Koenig2007}. While this approximation is valid for the related monolayer QSHI bismuthene \cite{Reis2017}, the discrepancy suggests already at this point a non-linear edge dispersion at the boundary of indenene \cite{Bieniek2023}.}

\begin{figure*}[t!]
 \centering
	\includegraphics[width=1\textwidth]{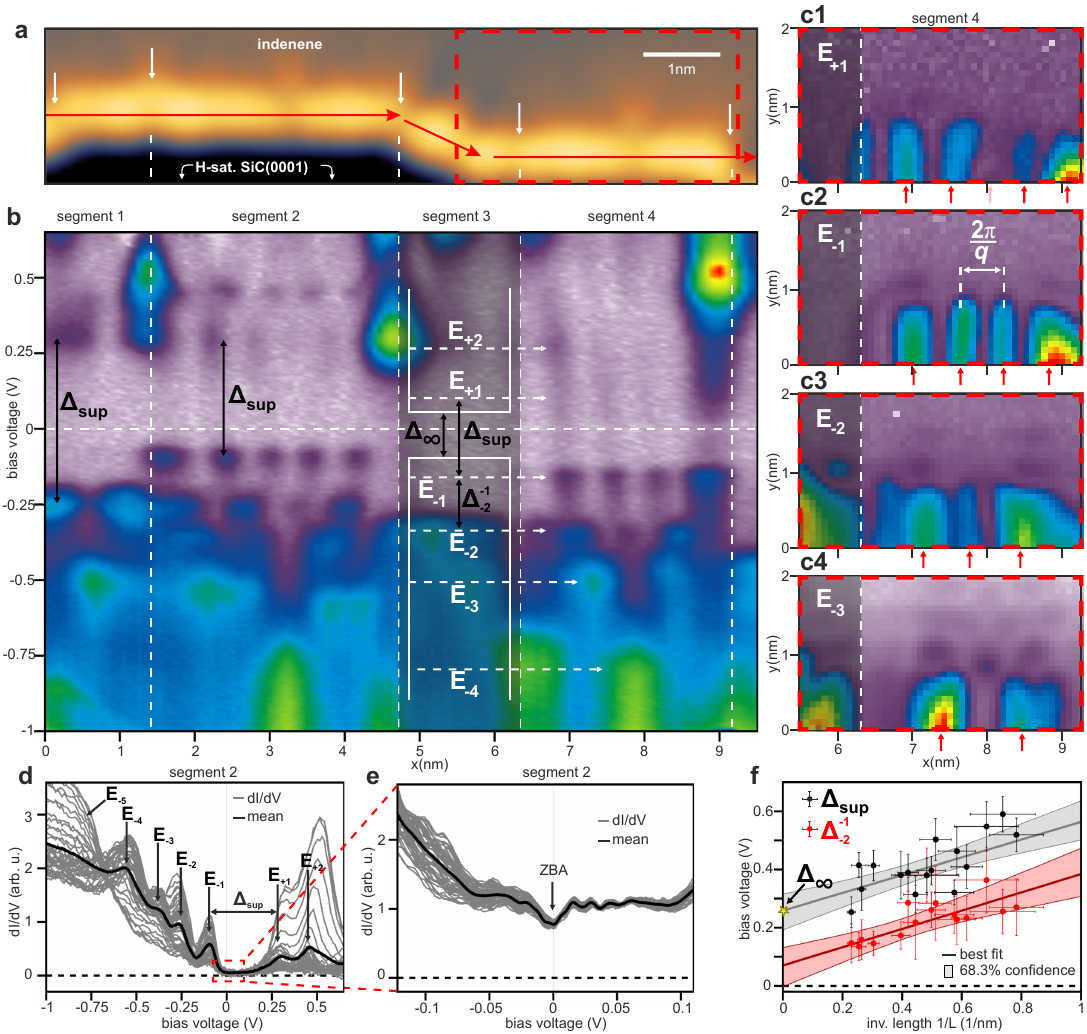}
\caption{
\textbf{Backscattering in A-terminated indenene edge states.}
\textbf{a} Indenene edge topography on H-sat. SiC, acquired at low corrugation settings (I$_{\text{T}}= 500 \,$pA, V$_{\text{bias}}= -1 \,$V). Defects and kinks in the edge (white arrows) indicate scattering centers that separate four segments, with segments 1, 2, and 4 being of flat A type, see Supplementary Discussion I and III.
\textbf{b} Line scan of d$I$/d$V$($E,x$) taken along the red path in \textbf{a} showing typical QPI-related d$I$/d$V$ modulations confined to flat edge A segments as well as variations in $\Delta_{\text{sup}}$ among them. 
%
\textbf{c} Spatially resolved d$I$/d$V$($x,y$) maps of segment 4 (dashed red rectangle in \textbf{a}) taken at energies E$_{+1}$ to E$_{-3}$ (labeled in \textbf{b}) and demonstrating edge localization of the respective d$I$/d$V$ peaks (red arrows) that are spatially separated by $2\pi/q$ (I$_{\text{T}}= 600 \,$pA, V$_{\text{bias}}= -1 \,$V).
%
\textbf{d} Cumulative plot of segment 2 d$I$/d$V$($E$) curves showing d$I$/d$V$ peaks E$_{-5}$ to E$_{+2}$ as well as a energy region within $\Delta_{\text{sup}}$ of QPI suppression.
\textbf{e} Separate zoom measurement of d$I$/d$V$($E$) in segment 2 demonstrating absence of spatial d$I$/d$V$-modulation except for a mild ZBA as well as finite differential conductance in $\Delta_{\text{sup}}$ (I$_{\text{T}}= 250 \,$pA, V$_{\text{bias}}= 0.1 \,$V). 
%
\textbf{f} Inverse length $1/L$-dependence of the topmost level spacing $\Delta_{\text{sup}}$ and $\Delta^{-1}_{-2}$ in a quantum well picture, as indicated in the inset of \textbf{b}. Assuming linear $E(q)$ dependence \textcolor{black}{for these energies}, we fit a $\Delta = \Delta_{\infty} + \hbar v_{F}^e\pi/L$ dependence to the data yielding $\Delta_{\infty} = (0.26\pm0.06)$eV for $\Delta_{\text{sup}}$.
}
\label{Fig:3}
\end{figure*}

\subsection{Backscattering of indenene edge states}

Having demonstrated the existence of metallic edge modes by STS, let us now turn to their robustness with respect to elastic backscattering. A representative STM topography scan comprising one zigzag and three flat edge segments, interrupted by \textcolor{black}{edge imperfections} (white arrows), is shown in Fig.~\ref{Fig:3}a. An STS line scan along the edge, \textit{i.e.}, along the red path outlined in Fig.~\ref{Fig:3}a, is shown in Fig.~\ref{Fig:3}b. 

\textcolor{black}{The scan reveals characteristic modulations of the LDOS, that are spatially confined to individual edge
segments of length L. This characteristic particle-in-a-box
QPI pattern is a clear indicator of edge modes being
reflected back and forth between defects \cite{Drozdov2014,Seo2010,Stuehler2022},
stabilizing LDOS modulations with wavelengths $2\pi/q$, that submit to the individual
resonator condition of each segment.
The QPI-active energies significantly exceed the indenene bulk band gap, which is related to the non-linear $\mathcal{S}$-shaped edge state dispersion, as we will see later, and similarly predicted for related materials \cite{Pulkin2020}.
This is further supported by strong variations in the level separations $\Delta^{i}_{j} = E_{i} - E_{j}$ across the wide bandwidth.
However, near the bulk gap, we identify almost equidistant level spacings $\Delta^{-1}_{-2}$ and $\Delta^{-2}_{-3}$, evidencing a linear section of the $\mathcal{S}$-shaped edge band dispersion, in which the modes are equally spaced by $\Delta=\hbar v_{F}^e\pi/L$, with $v_{F}^e$ being the Fermi velocity of the contributing edge mode.} 

\textcolor{black}{Their spatial extension into the bulk is shown in Fig.~\ref{Fig:3}c for selected energy modes E$_{+1}$ to E$_{-3}$ of segment 4, demonstrating that the corresponding LDOS modulations localize within the decay length $\xi \sim 3.6 $\;\AA\;away from the edge and thus are clearly related to the edge modes of indenene.}

Inspecting the $dI/dV$ line spectra of segment 2 in Fig.~\ref{Fig:3}b more closely in Fig.~\ref{Fig:3}d, we further notice the expression of a characteristic energy region $\Delta_{\text{sup}}$ between E$_{+1}$ and E$_{-1}$ where d$I$/d$V$ is suppressed. While the size of $\Delta_{\text{sup}}$ varies strongly from segment to segment, it always exceeds the size of indenene's band gap $E_{\text{gap}}\sim 120$\;meV as well as the equidistant energy spacing $\Delta^{-1}_{-2}$ of the adjacent linear band section. 
Plotting $\Delta_{\text{sup}}$ and $\Delta^{-1}_{-2}$ as a function of $1/L$ for a variety of flat edges of different lengths in Fig.~\ref{Fig:3}f, we find $\Delta_{\text{sup}}$ to be universally offset by $\Delta_{\infty} =(0.26\,\pm0.06)$eV with respect to the expected particle-in-a-box spacing $\Delta^{-1}_{-2} =  \hbar v_{\text{F}}^e\pi/L$, even in the limit of large segment lengths $L\rightarrow \infty$ where $\Delta^{-1}_{-2} \rightarrow 0$ is supposed to vanish. This directly rules out a dynamical Coulomb blockade as origin for the $dI/dV$-suppression within $\Delta_{\text{sup}}$ (see Supplementary Discussion IV) \cite{Jolie2019}.

\textcolor{black}{Close-up $dI/dV$ measurements in Fig.~\ref{Fig:3}d resolve the LDOS within $\Delta_{\text{sup}}$ to remain always finite, thus clearly metallic, but essentially featureless except for a zero bias anomaly (ZBA), which suggests second-order effects that will be explored in future studies \cite{Stuehler2020}.}
Notably, we find no signs related to scattering derived QPI within $\Delta_{\text{sup}}$. We thus conclude the presence of metallic edge states within $\Delta_{\text{sup}}$ 
that are resilient to quasiparticle backscattering, as expected for an $n=1$ QSHI.

\subsection{Discussion - QPI and edge band dispersion}
\begin{figure*}[t!]
 \centering
	\includegraphics[width=1\textwidth]{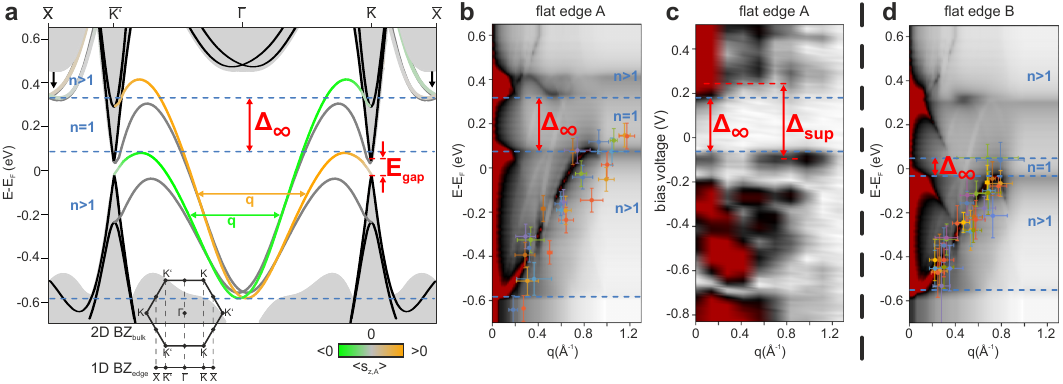}
\caption{\textbf{Edge bands and QPI pattern}
\textbf{a} Tight binding slab calculations (gray) of indenene flat edges with flat edge A localized bands color-coded according to their $\langle s_z \rangle$ character. 
For clarity, selected bulk bands (black) indicate valence and conduction band onsets with the bulk band gap $E_{\text{gap}}$ marked in red.
Further indicated for flat edge A are the number of Kramers pairs (blue) in a given energy interval as well as exemplary TRS-allowed backscattering momenta transfers $q$.
\textbf{b} Energy-dependent momentum JDOS $E(q)$ of $q$ involved in scattering processes that are facilitated by the dispersion of flat edge A, for details see Supplementary Discussion V.
Data points are extracted from experimentally observed resonator oscillations (see $2\pi/q$ in Fig.~\ref{Fig:3}c) and offset by 0.2$\,$eV to compensate for the experimental n-doping, see Method section. Each color represents one edge segment.
\textbf{c} Modulus of fast Fourier transform of segment 2 in Fig.~3\textbf{b} showing $q$ related spectral weight, QPI suppression (dashed lines) and onset of QPI at positive bias voltages.
\textbf{d} Calculation of momentum JDOS $E(q)$ associated with flat edge B bands.
Experimental data points are extracted from flat edge B segments similar to \textbf{b}.
}
\label{Fig:4}
\end{figure*}

Let us now reconcile the edge mode QPI observed in STS with theoretical expectations. For this purpose, we employ density functional theory (DFT) calculations of bulk indenene \cite{Bauernfeind2021,Eck2022} to construct tight-binding (TB) slabs for all edge types and project the resulting band dispersions onto their respective 1D Brillouin zones. Results for the indenene flat edge are shown in Fig.~\ref{Fig:4}a, where boundary modes localized at flat edge A are color-coded according to their individual spin character $\langle s_z \rangle$. Modes that localize at the opposite edge, \textit{i.e.}, flat edge B, are shown in dark gray for reference. Detailed calculations of spin and orbital character of all edge terminations are presented in Supplementary Discussion V.

The metallic states at flat edge A arise from two strongly dispersing $\mathcal{S}$-shaped bands with opposite and nearly constant $\langle s_z \rangle$ spin polarization. This overall deviation from the more common linear band scenario (cf. Fig. \ref{Fig:1}a) is ascribed to the multi-orbital p$_{x,y}$ and p$_z$ character of the edge states, 
along with the local potential \textcolor{black}{induced by the edge termination,} which shifts the TRS protected crossing point at $\Gamma$ towards higher binding energies $\sim \,-0.6\,$eV \cite{Hasan2010,Eck2022}.
Matching their spin character, the edge states connect the \textit{first} valence with the \textit{second} conduction band of the projected 2D band structure (see black bands) \cite{Eck2022}. Additionally, a set of Rashba-split free electron parabolas centered at the $\overline{X}$ points of the edge BZ (black arrows) are attributed to the edge and exhibit a dominant $\langle s_y \rangle$ spin polarization (see Supplementary Discussion V).

Clearly, the $\mathcal{S}$-shaped edge dispersion of flat edge A exhibits the qualitative scenario discussed in the context of Fig.~\ref{Fig:1}b, 
\textcolor{black}{namely an energy interval with only a single ($n=1$) Kramers pair, and self-overlapping sections of the dispersion with $n>1$ Kramers pairs, respectively.}
As described above, the latter allows for \textit{intra-band} backscattering via characteristic momentum transfer $q$ which connects two different Kramers pairs, as exemplary marked in Fig.~\ref{Fig:4}a. To identify all available \textcolor{black}{TRS-preserving} scattering channels $q$ consistent with the edge dispersion, we apply the T-matrix formalism to calculate the explicit energy dependent \textcolor{black}{momentum joint density of states (JDOS) $E(q)$ (Fig.~\ref{Fig:4}b), representing an autocorrelation function of the edge bands weighted by scattering matrix elements (see Supplementary Discussion V).}

Clearly, the favorable scattering channels $q$ disperse along continuous $E(q)$ \textcolor{black}{branches}. Within the $n=1$ energy region where scattering is strictly prohibited by TRS, they exhibit a gap of size $\Delta_{\infty}\sim0.25$\;eV, in excellent agreement with the $\Delta_{\infty}=(0.26\,\pm0.06)$\;eV that was found in the experimental data of Fig. \ref{Fig:3}f. Further, extracting the energies $E$ and wavelengths $2\pi/q$ from LDOS modulations in a variety of $dI/dV$ line-scans (e.g. Fig.~\ref{Fig:3}c top) and plotting the resulting $E(q)$ data points on top of the calculated momentum JDOS $E(q)$ in Fig.~\ref{Fig:4}b, we also find excellent match of theory with experiment. \textcolor{black}{The same applies} for the in-gap Fermi velocity $v_F^e$, which we extract both from the slope of the linear region in $E(q)$, as well as in the slope of the data in Fig. \ref{Fig:3}f, and find $\hbar v_{F}^e =(0.9\pm 0.1)$\,eV\AA$ $ and $\hbar v_{F}^e =(1.0\pm 0.3)$\,eV\AA$ $, respectively.
Finally, we directly calculate the Fourier transform of the d$I/$d$V$ line-scan  for segment 2 and plot the result in Fig.~\ref{Fig:4}c for comparison. Despite the spatial confinement of the segment, which produces diffuse spectral features along the $q$ axis of this plot, its overall resemblance to the calculation in Fig.~\ref{Fig:4}b is convincing.

For completeness, we underline this remarkable agreement between experimentally observed and theoretically predicted backscattering further by comparing analogous calculations for flat edge B with experimentally determined $E(q)$ data in Fig.~\ref{Fig:4}d. Again, the TRS protected $n=1$ region is marked by an -- albeit smaller -- gap where electron backscattering is prohibited. In contrast to flat edge A, however, the connection of \textit{second} valence and \textit{first} conduction bulk band deforms the edge bands near the projected valley momenta $\overline{K}/\overline{K'}$ in such a way, that multiple shorter momentum transfers $q$ are allowed. As the corresponding wavelengths $2\pi/q$ are much longer than the segment lengths available experimentally, their extraction from $dI/dV$ lies beyond our capabilities. We note that similar arguments apply to the zigzag edges (see Supplementary Discussion VI).

The spectral properties of indenene edge states laid out in this work reveal elastic single-particle backscattering among different sets of Kramers pairs, driven by the non-linearity of their specific edge dispersions.
\textcolor{black}{Consequently, in edge transport measurements of indenene, the presence of disordered defects as scattering centers suppresses the two terminal conductance to its minimum quantization value of 2e$^2$/h \cite{Wada2011,Li2014,Bieniek2023,Martin1992,Fisher1981}, rather than the purely ballistic value of $n\times$2e$^2$/h expected for $n$ Kramers pairs.
This conclusion is of general importance for transport measurements in all QSHIs featuring nonlinear edge band dispersions.
Notably, it is especially relevant for WTe$_2$, where band calculations indicate $n=3$ Kramers pairs \cite{Zheng2016}, while experimental measurements reveal 2e$^2$/h \cite{Wu2018}. This suggests that two Kramers pairs are eliminated by disorder \cite{Wada2011,Li2014,Bieniek2023,Martin1992,Fisher1981}, precisely as observed in this work.
}

%% file: Methods.tex
\section*{Methods}

\noindent \textbf{Sample preparation.}--- Indenene was grown on N-doped 4H-SiC substrates ($12\,\text{mm}\times2.5\,\text{mm}$) with a resistivity of $(0.013)\,\Omega\text{cm}$ (Fig.~2) and $(0.015 - 0.03)\,\Omega\text{cm}$ (Fig.~3,4). 
Atomically flat 4H-SiC surfaces were prepared by a dry-etching technique that saturates the silicon dangling bonds with hydrogen and stabilizes the ($1\times1$) SiC(0001) surface \cite{Glass2016}. 
Except for data shown in Fig.~2, substrates with a higher dopant concentration are used for practical reasons when performing STM near H-sat. SiC regions, thereby shifting the indenene band gap to negative bias voltages, see Supplementary Discussion II.

Indenene films were grown by a simple two step process described elsewhere \cite{Erhardt2022}. In order to increase the indenene edge density on the surface (Fig.~\ref{Fig:2}b), we reduced the hydrogen-desoption duration (at 600$^{\circ}$~C) to 3$\,$min as well as indium accumulation (at 420$^{\circ}$~C) to 25$\,$min. 
This preserves $\sim 10\%$ of the H-sat. SiC surface within the otherwise closed indium film.
The latter is transformed to indenene in the second growth step by postannealing at 480$^{\circ}$~C for 15$\,$min \cite{Erhardt2022}.  
The substrate temperature was measured with a pyrometer (Keller, detection range 1.1–1.7 $\mu$m, emissivity $\epsilon = 85\%$) sensitive to a temperature range of 250-2000$^{\circ}$~C.
Indium of 99.9999$\%$ purity was evaporated from a Knudsen cell held at 770$^{\circ}$~C creating an indium flux of $\sim 0.05$\AA/s on the substrate surface as measured with a quartz crystal microbalance.

\noindent \textbf{ARPES measurements} were performed in our home-lab setup equipped with a hemispherical analyzer (PHOIBOS 100), a He-VUV lamp ($\mu$SIRIUS, 21.2$\,$eV), and a 6-axis manipulator capable of LHe-cooling to 20\,K. ARPES data shown in Fig.~2c were recorded at 20\;K and a base pressure of $<10^{-10}\,$mbar. To compensate geometric asymmetry of the experiment, each energy distribution curve  is normalized to its integral intensity. Differential pumping of the He-VUV-lamp kept the base pressure below $10^{-9}\,$mbar during the ARPES measurements.

\noindent \textbf{STM measurements.} STM data were acquired at 4.7$\,$K and a base pressure lower than 5$\times10^{-11}\,$mbar (Omicron low-temperature LT STM) using a chemically etched W-tip that was characterized by imaging the Ag(111) surface state. 
Point spectroscopy dI/dV curves were recorded using a standard lock-in technique with a modulation frequency of 787$\,$Hz and modulation voltage of V$_{\text{rms}}$=7$\,$mV (Fig.~2), V$_{\text{rms}}$=1$\,$mV (Fig.~3e) and V$_{\text{rms}}$=10$\,$mV (Fig.~3b,c,d and Fig.~4b,c,d).

\textbf{Inverse length analysis} in the quantum well regime (Fig.~3f) is performed by fitting Gaussian profiles to isoenergetic LDOS maxima, allowing us to determine the corresponding bias voltage. The error associated with each peak is given by the Gaussian standard deviation.
The segment length $L$ is evaluated based on topography scans, with its error estimated accordingly. 
Fits to data shown in Fig.~3f are obtained using linear regression. 

\textbf{$E(q)$-analysis} presented in Fig.~4b,d is determined from the separation between isoenergetic LDOS maxima on the x-axis (indicated in Fig.~3c2), each fitted with Gaussian profiles yielding their x-axis position.
The wavelength $2\pi/q$ is then calculated as the average x-separation of neighboring LDOS peaks.
The errors for each maximum in the x-direction and bias voltage correspond to their respective Gaussian widths and are propagated to calculate the standard error in $2\pi/q$.
For the lowest LDOS maximum, $2\pi/q$ has to be estimated from the quantum well length and is therefore less precise. 
For more details see Supplementary Discussion III.

\textbf{DFT and tight binding calculations of indenene.}
First-principles calculations are performed using Density Functional Theory (DFT) as implemented in the Vienna Ab-initio Simulation Package (VASP) \cite{VASP1,VASP2,PAW}. The exchange-correlation potentials used are the HSE06 hybrid functional \cite{HSE06,Heyd2003} and PBE \cite{Perdew1996,Perdew2008} for bulk and slab calculation respectively, in a non-collinear magnetic moment configuration with self-consistently calculated SOC \cite{SOC_VASP} where specified. The energy cutoff for the plane-wave expansion is set to $500~\mathrm{eV}$, while the Brillouin zone is sampled on a $12 \times 12 \times 1$ regular mesh.\\
The bulk structure considered in DFT calculations is the $1 \times 1$ reconstruction of a Si-terminated four-layer SiC($0001$) substrate with indium atoms in T$_1$ positions. The In-SiC distance is $2.68~\mathrm{\AA}$ after converging the forces to within $0.005 \mathrm{eV/\AA}$. A vacuum region in the $z$-direction of $25 \mathrm{\AA}$ is used to disentangle periodic replicas.\\
Tight binding and $E(q)$ calculations are based on these DFT results and summarized in more detail in Supplementary Discussion V.